\def\year{2018}\relax
\documentclass[letterpaper]{article} 
\usepackage{aaai19}  
\usepackage{times}  
\usepackage{helvet}  
\usepackage{courier}  
\usepackage{url}  
\usepackage{graphicx}  

\begin{document}
%
\title{Consistency of Forecasts for the U.S. House of Representatives}
\author{Henry Bendekgey\\
Pomona College\\
Claremont, CA\\
henry.bendekgey@pomona.edu}
\maketitle

\begin{abstract}
We consider the performance of the foremost academic House of Representatives forecasting models in the 2018 elections. In creating open-source implementations of these models, we outline key underlying assumptions. We find that although the results were unsurprising, they indicate a weakening of many traditional forecasting indicators.

\end{abstract}

\section{Introduction}

The results of the United States 2018 midterm elections are in, and with only one race left uncalled three weeks after election day, the Democrats have gained 39 or 40 seats in the House of Representatives. Although this outcome---the so-called ``blue wave"---is consistent with most predictions leading up to the election, the individual forecasting indicators pointed in wildly different directions as November 6 approached.

Fivethirtyeight's final prediction of a net pickup of 39 seats ended up being right on target \cite{538}. The Economist predicted a net change of 36 seats \cite{Economist}. By synthesizing all of the indicators, they made robust predictions that held up to reality. Academic models, on the other hand, varied much more wildly in their predictions, from a predicted change of 27 seats to a predicted change of 44 seats.

In the October 2014 issue of \emph{PS: Political Science and Politics}, five teams presented forecasts for the U.S. House of Representatives midterm elections. Four of these forecasts were accurate to within four seats, and the last was off by nine \cite{Recap2014}. One team noted that, given the uncertainty associated with their prediction, to have gotten a result so close to their forecast required some degree of ``luck" \cite{Recap2014}. 

Two years later, the recap of the 2016 forecasts wasn't as celebratory \cite{Recap2016}. Forecasts predicted Democratic gains in the house that never materialized. However, predictions were not wildly off; many predicted the Presidential national vote with high precision, and house predictions were mostly within standard prediction errors. The error was consistent across the forecasts, with predictions favorable to the Democrats, which might have accidentally conveyed a confidence in these results that didn't exist in any of the models individually.

In 2018, the various tools used to make predictions were pointing in different directions, some towards a modest Democratic gain and some towards a huge blue wave. Four of the models presented in the October 2014 issue of \emph{PS} appeared again in the 2018 issue. We created an open-source implementation of each of these models in advance of their publication, to examine the underlying assumptions that might cause some of them to perform better than others.

Our replications of these models were extremely consistent with the results published in October (See Table \ref{compare}). It is a good sign for the validity of these models that they can be independently reproduced to a great deal of accuracy. The notable exception is the Seat-in-Trouble model, which had a significant change in methodology for the 2018 election.

\begin{table*}
    \begin{center}
    \begin{tabular}{|l|c|c|}
        \hline 
        Model & Reproduced Predicted Seat Change   & Published Predicted Seat Change      \\ \hline
        Generic Ballot           	 & -32 & -30 \\
        National Polls District Info & -28 & -27 \\
        Seats-in-Trouble             & -68 & -44 \\
        Structure-X.                 & -43 & -44 \\ \hline
    \end{tabular}
    \caption{\textmd{Predicted House seat change for Republican party in reproduced and official forecasts. Actual result is expected to be -39 or -40 based on California's 21st district.}}
    \label{compare}
    \end{center}
\end{table*}

In this paper we examine the stated and unstated assumptions of each model. We note that the parsimonious models based on historical precedent under-explained the blue wave, indicating that either 2018 was an exception, or that the effect of these national conditions are changing.

\section{Background}
Midterm election models largely fall into three camps: Synthesizers, Structuralists, and Experts \cite{Camps}. Synthesizers are the most common, or at least the most pervasive to nonacademics: they combine structural knowledge with polling data to provide constantly shifting predictions that swing with the daily polls. This is what Fivethirtyeight \cite{538}, the Economist \cite{Economist}, and other popular websites are doing.

However, a forecast produced on the day before the election is worth little compared to one produced months in advance. An early enough forecast can help election stakeholders make decisions about campaign strategies and the allocation of resources. Academic literature has shown that it is possible to make a prediction long in advance \cite{Gelman}. 

Synthesizing models promote obsession with the daily swings of polls, which, although exciting, ultimately do not have much effect on the outcome. Further, the volume of predictions produced can pose a problem to a casual observer: for example, consider that close to election day the probability of some outcome remains steady at 80\%. A casual observer could be tempted to believe that those daily probabilities reinforce each other, and round up that value, believing the outcome to be more certain than just 80\%. In truth, those predictions are highly correlated, and the outcome is hardly certain. 

Finally, although Fivethirtyeight promotes explanations of the underlying math in their model, how they chose their parameters and the model's exact workings remain a black box for proprietary reasons. Conversely, the advantage of academic predictions is that they can be replicated and augmented, making projects like this one possible. 

The second camp of forecasters, structuralists, try to predict using situational factors that are rooted in political theory \cite{Camps}. For example, if the economy is doing poorly, forecasters would expect voters to swing away from the party currently in control of the government. Structuralists use data like economic performance in their forecasts.

All of the forecasts we will be examining use post-World War II election data to make their predictions, defining the elections in this time period to be modern elections on which we can regress to make conclusions about today. This makes 2018 the 19th modern midterm election. With so few data points, it would be easy to believe there are trends where there is only coincidence, overfitting a model to the data from previous elections. Thus predictors need to be rooted in theory, and parsimony is highly valued. In that way, election forecasts serve a dual purpose: not only to make predictions about the future, but to reinforce or undermine political theories about what causes a party to gain or lose influence. This is especially important this election, where different predictors were pointing towards different outcomes, so the ultimate result can speak to the relative importance of these predictors.

The disadvantage of such models is that they rest on the underlying assumptions of linear regression; they assume that the effect of each national condition on voting patterns has remained constant since World War II. Any deviation from these standards would result in a poorly-predicted election.

The final camp, known as experts, use expert opinions months in advance to predict election results. Organizations like \emph{Cook Political Report} produce seat ratings in the months before elections, rating individual races as ``Safe", ``Likely", or ``Leaning" towards a particular party, or ``Toss-up" \cite{Cook}. These ratings tend to be well-correlated with the probability that a seat flips \cite{Campbell2018}, so experts make predictions based on these ratings. These have historically been extremely accurate, but in 2016 they performed notably poorly.

We will discuss the four forecasts published in both the 2014 October issue and 2018 October issue of \emph{PS: Political Science and Politics}.

\section{Forecasts}
\subsection{Generic Ballot Model}

\begin{figure}
\begin{center}
	\includegraphics[width=\linewidth]{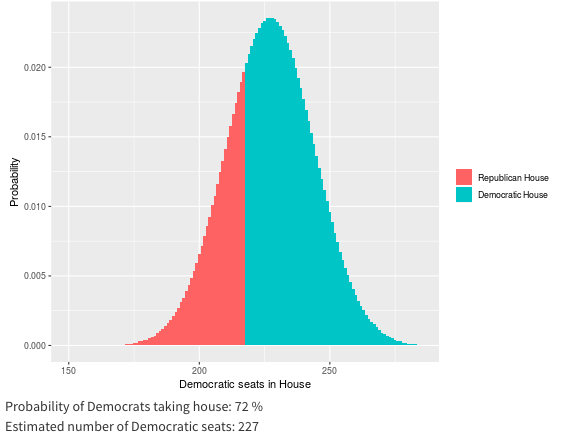}	
\end{center}
\caption{\textmd{House seat probability distribution from reconstructed Generic Ballot Model. Blue fill indicates a Democratic-controlled House, while red indicates a Republican-controlled House.}}
\end{figure}

Professor Alan Abramowitz proposes a simple structuralist model for house and senate prediction. To predict the net change in Republican-held seats in a midterm election year, we need three pieces of information: which party currently holds the Presidency, the generic ballot margin in early September, and the number of seats Republicans currently hold \cite{Abramowitz2018}.

Each one of these predictors is heavily rooted in theory. The party which holds the Presidency tends to lose seats during the midterms. In fact, 1998 and 2002 were the only midterm election years since World War II where the incumbent party did not lose seats in the House. In addition, parties that hold more seats will tend to lose more; this is called an exposure model \cite{Campbell2014}.

The generic ballot margin refers to the net approval of Republicans in so-called generic ballot polls, national polls in which registered or likely voters are asked if they will vote for the Democrat or Republican in their district. This margin is clearly a reflection of voter sentiment, and is pertinent to how voters will act in November.

The primary problem with the reproduction of this model is the consistency of generic ballot data. One of Abramowitz's predictors is the generic ballot margin in early September. We interpreted this to mean the average margin in polls taken 60 to 90 days before the election, because the exact timeframe is not specified. In many of the earlier elections being regressed on, this value reflects a single poll, usually by Gallup. In recent years, polling has proliferated massively. Reuters/Ipsos releases a new poll every day. We follow the methodology of Professors Bafumi, Erikson and Wlezien and include only live polls, including those taken over the telephone by real people \cite{Bafumi2014}. This is not because we believe live polls to be more accurate than robotic polls. Rather, this is an attempt to maximize consistency across the elections being regressed on. Thus we exclude polls taken online, robotically, or by IVR (Interactive Voice Response). 

However, this poses an existential threat to forecasting models that try to use historical polling data: technology is changing, as is the way we interact with it. The assumption that the relationship between polling results and voter sentiment has remained constant over these technological changes through the past 70 years is hard to defend and even harder to prove.

Ultimately, we produce a forecast that differs from Abramowitz's by 2 seats, a difference that can most likely be attributed to the use of different polling data.

This model ultimately under-predicted the Democratic gains in 2018, indicating the fervor of Democratic voters was not fully captured by the generic ballot polling results.

\subsection{National Polls and District Info Model}

Bafumi, Erikson, and Wlezien propose a structuralist model for forecasting US house midterm elections based on data available by early July \cite{Bafumi2018}. The first step in the NPDI model is to predict the national house vote for the upcoming midterm election. This is done using only two predictor variables: the party of the current president, and the average Democratic share of generic ballot results between 180 and 121 days before the election.

The response variance in their linear regression is the Democratic share of the two-party national House vote, measured in percentage points away from 50. However, in their model they do not force the intercept to 0. The empirical intercept value of -0.14 reflects a belief that independent of other factors, Democrats will outperform Republicans. There isn't any theory to defend this, only the data, and concerns about overfitting indicate we might want to remove this term. Thus in the replication of the model, we force the intercept to 0.

Bafumi et al. predict Democrats to win 53.6\% of the national house vote, which matches our model's replication of 53.5\%. At this time, the true Democratic two-party share of the national house vote is 54.0\%.

Another important question to address at this point is why we care about the national vote. Ultimately, we are interested in the swing of national House vote from 2016 to 2018. In order to model the covariance of House races, Bafumi et al simulate the election by first picking a value for the national swing using a prediction interval from their linear model. They then assume that on average, districts move uniformly to the left by that swing, and simulate uncertainty in each race's result.

This is done repeatedly, simulating thousands of elections. The proportion of these elections in which a certain event occurs is taken to be the probability of that event occurring.

\begin{figure}
\begin{center}
	\includegraphics[width=\linewidth]{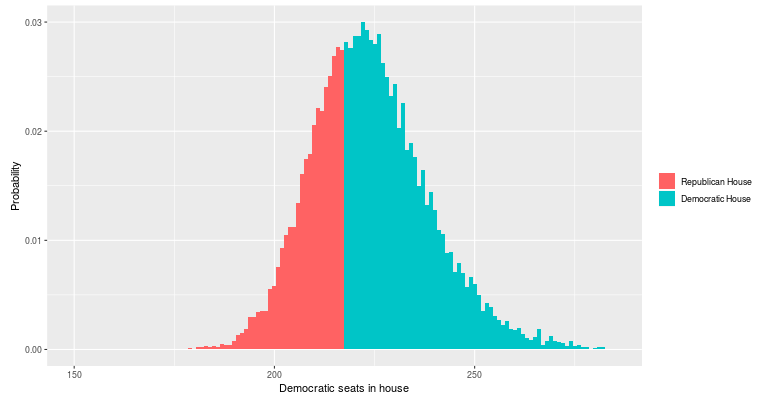}	
\end{center}
\caption{\textmd{Results from 10,000 simulations of reconstructed National Poll District Info Model.}}
\end{figure}

There are a few things to unpack here. In truth, we are not interested in the swing of the national House vote. We are interested in the mean swing in contested districts. If we knew that value, then the shift of all contested races by that amount would create an unbiased estimator for the actual results. But when a district is uncontested, it puts bias on the national House vote that does not reflect the change in national conditions. This is particularly exacerbated by top-two primary states like California, where some races don't include a single Republican. An increase in uncontested races will therefore be disproportionately represented in the national vote percentage.

In addition, the total swing in contested districts and the mean swing in contested districts are not the same. For example, if there is a negative correlation between percentage Democratic vote in a district and the total number of people that vote, we would expect that summing all the votes in those districts and calculating percentage Democratic vote would look worse for Democrats than just taking the mean Democratic share across those districts, because the latter weighs all districts evenly. Thus if one party's turnout is dampened or increased more than the other party's from 2016 to 2018, that introduces bias into the model.

In their 2014 model, Bafumi et al argue that in midterm election years Democrats are particularly bad at turnout and the Democratic mean district advantage grows in size \cite{Bafumi2014}. However, after observing the results of the 2014 election, they found this effect to be not as consistent as they proposed in 2014, and abandoned that adjustment in their 2018 model.

How to take into account the discrepancy between national vote and mean competitive district vote remains an open problem. However, we independently came to the same conclusion, and simply used national vote as a proxy for mean competitive district vote, as there is not enough reason to believe that prediction would be more likely to be wrong in one direction than another.

Finally, the mean swing is used to predict each individual race based on how it behaved in 2016. Bafumi et al propose only dividing races up into those with incumbents and those with open seats. Any race that was uncontested in the previous election is conceded immediately. This will rarely pose a problem, because it is unlikely a seat will go from uncontested to competitive in 2 years.

However, this does pose a potential danger, because that does happen. For example, PA-18 was not contested by Democrats in 2016, but Democrats actually won it in the early 2018 special election, although due to redistricting that district no longer exists. It happened again on November 6 when TX-32 was won by Democrats despite not fielding a candidate in 2016.   

The uncertainty in incumbent races is found to be smaller than in open seats, which matches our expectation. Freshman surge, the gaining of incumbency advantage for a freshman representative, is further calculated and included in the model.

In this model, we are predicting how districts behave based on how they have behaved in 2016. However, the question of what that even means is non-trivial. For incumbent races, do we use how they voted for representative, or how they voted for president? In open races, we only use presidential performance, because the candidate voted on in 2016 and 2018 are different. But in incumbent races, we want to use both. Because these values are hugely collinear, the relative power of these predictors is difficult to calculate from the data. 

Further, this is a political question. In the lead-up to the 2018 election, a big question in the air was: is the GOP the party of Trump? Will Republicans perform better in districts they historically perform well in, or in the ones specifically Trump performed well in?

Despite a number of simplifications, we predicted the Republicans would lose 28 seats, while Bafumi et al. predict they would lose 27. This is a dramatic underestimate of the real results, which starkly contrasts with the accuracy of the national vote share.

What can we learn from the fact that the model accurately predicted the national vote, but not the seat swings? Plenty of districts swung more to the left than Bafumi et al. predicted, but elsewhere there must have been more Republican votes to compensate. The easiest explanation for this discrepancy is that the model of uniform district swing didn't hold: some demographics, either racial, gender, education level, or geographic, swung more to the left than average, handing more districts to the Democrats than Bafumi et al. expected. Other demographics, to compensate, must have swung left.

\subsection{Structure-X Model}

Michael Lewis-Beck and Charles Tien mix a structural model and an expert model to produce their forecast, giving it its name: Structure-X \cite{LewisBeck2018}. 

In this model, the change in incumbent party seats is predicted by three variables: the percent change in Real Disposable Income (RDI) in the first two quarters of the election year, the approval rating of the President in June of the election year, and whether or not it is a midterm election. Each of these predictors is rooted in theory and points in the direction we'd expect: good economic growth is associated with pro-incumbent sentiment, as does Presidential approval. Incumbents tend to lose seats during midterm years, having done so in every election since 1948 other than 1998 and 2002, the latter being an exceptional case as the election immediately followed 9/11.

Although it makes sense that presidential approval predicts midterm results, it would also make sense to claim presidential disapproval is a good predictor. These two are highly correlated, and result in almost identical $R^2$ values. In this election, the choice of which to use makes little difference. Predicting using disapproval results in an extra swing of 1 seat towards Democrats.

\begin{figure}
\begin{center}
	\includegraphics[width=\linewidth]{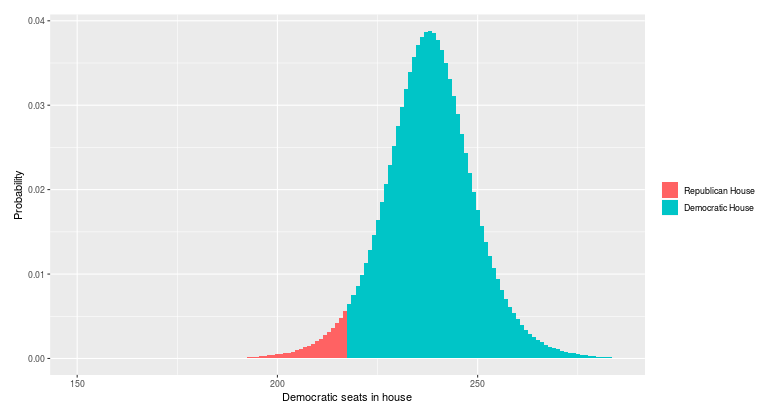}	
\end{center}
\caption{\textmd{House seat probability distribution from reconstructed Structure-X Model.}}
\end{figure}

Using only these structural cues gives an estimated Republican loss of 28 seats in our replication, and 30 seats in Lewis-Beck and Tien's.

This model doesn't take into account the current distribution of seats, which is highly correlated with the resulting seat change. To close this predictive gap, Lewis-Beck and Tien take the average of their prediction and \emph{Rothenburg}'s Seats-in-Play differential \cite{LewisBeck2014}. Rothenburg synthesizes predictive factors to rate individual races as ``Safe", ``Likely", or ``Leaning" towards a particular party, or ``Toss-up". For the 2018 election, Rothenburg has been succeeded by \emph{Inside Elections} \cite{IE}. 

Why an average of these two values? Should one be weighted more than another? The problem with questions like these is that Rothenburg has only been producing forecasts since 2006, so any attempt to regress on how to combine these two predictors on only 6 data points will yield noisy results. Thus we follow the simple methodology of an equally weighted average. This yields good predictions for the past few elections, but with so few data points, it's hard to know if this is a coincidence. More elections might reveal that a differently weighted average will provide better results.

It is worth noting that the structural half of this model underestimated Democratic gains and Expert models overestimated them, but together they produced a slight overestimate.

\subsection{Seats-in-Trouble Model}

\begin{figure}
\begin{center}
	\includegraphics[width=\linewidth]{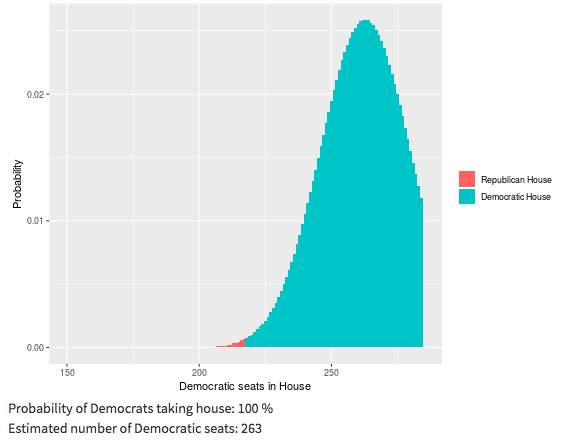}	
\end{center}
\caption{\textmd{House seat probability distribution from reconstructed Seats-in-Trouble Model.}}
\end{figure}

The Seats-in-Trouble model is an simple model built on a single predictor \cite{Campbell2018}. Using the paradigm of an exposure model introduced above, this model uses data from \emph{Cook Political Report}'s seat ratings to predict election results. The idea behind this model is that a party with more seats considered ``in trouble" will lose more seats, on average. However, in 2016 and before, ``in trouble" was defined to mean the number of seats a party controls that are rated as leaning towards them or worse \cite{Campbell2014}. The number of Democratic seats in play minus the number of Republican seats in play, called the net seats in play, is the only predictor for this model.

One large disadvantage for this model is that we don't understand what is motivating the predictions, because Cook does not publish their methodology. This model is purely predictive.

Using this methodology, we would predict a net change of seats of 68 towards Democrats, a massive wave. In truth, expert models predicted the 63-seat wave of 2014 better than any other type of models \cite{LewisBeck2014}. One concern, however, is that in this election, there are twice as many ``lean Republican"s than ``lean Democrats"s. Thus by reducing the distribution across multiple categories into a single predictor we lose information, and we might expect the real result to be to the right of this prediction.

Campbell argues that after the poor performance of the model in 2016, he re-evaluated, and redefined ``in trouble" to mean toss-up or worse. The failure of the model in 2016 can also be attributed, however, to experts predicting much better Democratic performance than materialized. 

Considering the breadth of factors that likely go into the expert predictions, it is concerning to consider two interpretations of the data, especially if the interpretation is chosen after seeing the data. Without enough solid reasoning to choose one model over another except for a difference in $R^2$ of 0.05 on such a small dataset, the choice is arbitrary.

Both interpretations of the data overestimated the seat change, albeit ours much more drastically than Campbell's.

\section{Conclusions}

After the 2016 election, there was concern about the reliability of polls \cite{Recap2016}. However, in 2018 national factors such as the performance of the economy and presidential approval ratings universally underestimated the size of the blue wave. The district-to-district polls and special election results pointed towards a massive wave.

These results informed expert prognosticators which in turn informed expert models. These models overestimated Democratic gains. The structural models, on the other hand, underestimated Democratic gains. Thus the problem lies not in polling specifically, but in the danger of ascribing too much confidence to any one indicator. Taken altogether, they produce a robust forecast. 

The underlying mechanisms of these models raise questions about the validity of linear regression in this setting. As technology changes, the relationship between polling results and voter sentiment might be changing as well. Further, the universal underperformance of fundamentals in predicting this election points to their power potentially weakening. Despite Trump being only moderately unpopular according to approval and disapproval ratings, and despite the excellent performance of the economy, Republicans lost dozens of seats November 6. The circumstances that inform voter preferences change over time, and it is important we build models to reflect these changes.

\section{Reproducibility}

All of the code for these models can be found at\\ github.com/hbendekgey/house\_forecast\_2018. The models can be further examined and adjusted at \\https://midterms.shinyapps.io/2018/. Finally, the codebase of that website can further be found at\\
github.com/hbendekgey/midterms-website

Missing from this codebase is generic ballot polls, which were kindly shared to me by Professors Bafumi, Erikson, and Wlezien. 2018 polls can be found in Fivethirtyeight's Github.

\bibliographystyle{aaai}
\bibliography{paper.bib}

\end{document}